\documentclass{aa}
\usepackage{graphicx}
\usepackage{txfonts}
\usepackage{epsfig}
\usepackage{times}

\begin{document}

\title{The bolometric luminosity of type 2 AGN from extinction-corrected
  [OIII]: no evidence for Eddington-limited sources.}
   \author{A. Lamastra\inst{1}, S. Bianchi\inst{1},
     G. Matt\inst{1}, G. C. Perola\inst{1}, X. Barcons\inst{2},
     F. J. Carrera\inst{2}}
   \offprints{lamastra@fis.uniroma3.it}
   \institute{Dipartimento di Fisica ``E. Amaldi'', Universit\`a degli Studi
     Roma Tre, via della Vasca Navale 84, I-00146 Roma, Italy
   \and 
   Instituto de F\'\i sica de Cantabria (CSIC-UC), Avenida de los Castros, 39005
   Santander, Spain
}

   \date{Received ; Accepted }

   \abstract{There have been recent claims that a significant fraction of type
     2 AGN accrete close or even above the Eddington limit. In type 2 AGN the
     bolometric luminosity (L$_{b}$) is generally inferred
     from the [OIII] emission line luminosity (L$_{OIII}$). The key
     issue, in order to estimate the bolometric luminosity in these AGN, is therefore to
     know the bolometric correction to be applied to
     L$_{OIII}$. A complication arises from the fact that the observed
     L$_{OIII}$ is affected by extinction, likely due to dust within the
     narrow line region.
     The extinction-corrected [OIII] luminosity (L$^{c}_{OIII}$) is a better
     estimator of the nuclear luminosity than L$_{OIII}$. However, so far
     only the bolometric correction to be applied to the uncorrected L$_{OIII}$ has been evaluated.}   
 {This paper is devoted to estimate the bolometric correction
   C$_{OIII}$=L$_{b}$/L$^{c}_{OIII}$ in order to derive the Eddington ratios for the type 2 AGN in a sample of SDSS objects.}{We have collected from the literature 61
   sources with reliable estimate of both L$^{c}_{OIII}$ and X-ray
   luminosities (L$_{X}$)
   . To estimate
   C$_{OIII}$, we combined the observed correlation between L$^{c}_{OIII}$ and
    L$_{X}$ with the X-ray bolometric correction.}{We found, contrary to previous studies, a linear correlation
   between L$^{c}_{OIII}$ and L$_{X}$. We estimated C$_{OIII}$ using the
   luminosity-dependent X-ray bolometric correction of Marconi et al. (2004),
   and we found a mean value of C$_{OIII}$  in the luminosity ranges log L$_{OIII}$=38-40,
  40-42, and 42-44 of  87, 142 and 454 respectively. We used it to
     calculate the Eddington ratio distribution of type 2 SDSS AGN at
     0.3$<z<$0.4 and we found that these sources are not accreting near their
     Eddington limit, 
contrary to previous claims.}{}


\authorrunning{Lamastra et al.}

\titlerunning{The bolometric luminosity of type 2 AGN from [OIII]}

\maketitle

\section{Introduction}

Active galactic nuclei (AGN) are believed to be powered by accretion of gas
onto the black hole located at the centre of galaxies. 
The AGN bolometric luminosity (L$_{b}$) depends on the mass accretion rate and on the efficiency for the conversion of gravitational energy into radiation.
In this scenario, the luminosity produced by a black hole has a physical
limit, the Eddington limit
(L$_{Edd}$$\simeq$1.3$\times$10$^{38}$(M$_{BH}$/M$_{\odot}$) erg/s), at which
the radiation pressure due to the accretion of the infalling matter balances
the gravitational force of the black hole. 
The ratio between the bolometric and Eddington luminosity, $\lambda$=
L$_{b}$/L$_{Edd}$, is referred to as the Eddington ratio.

The knowledge of the Eddington ratio distribution and its evolution is very
important to constrain the predictions of theoretical models which link the evolution of the galaxies in the hierarchical clustering
scenario with the quasar evolution (e.g. Menci et al. 2003, 2004). In fact,
all  predictions regarding AGN, such as the luminosity function or the black hole mass
function of AGN relics, depends on the assumptions about this quantity. The
estimate of the Eddington ratio
 requires measurements of the black hole
mass, M$_{BH}$, and of the bolometric luminosity, L$_{b}$.

According to the unification model (Antonucci 1993), in the AGN classified as
type 1 the optical/UV continuum source and the surrounding broad emission line
region (BLR) are viewed without any substantial obscuration, while in type 2 AGN these regions suffer obscuration along the
observer's line of sight by intervening dusty material, most of which is
likely associated with a circumnuclear structure often referred to as the
torus. Even if some exceptions have been found (e.g. Bianchi et al. 2008),
the unification model can be safely assumed to provide a useful reference
frame for the vast majority of sources.

Progress in reverberation mapping of type 1 AGN (Peterson et al. 2004, 2005)
provided rather good evidence that the BLR size depends on the optical
continuum luminosity (Kaspi et al. 2000, 2005; Bentz et al. 2006). 
This size-luminosity relation, combined with the gas
velocity derived from the width of the emission lines, under the assumption
that the BLR clouds are in Keplerian motion around the black hole, has been
widely used to estimate the black hole masses, in large samples of type 1 AGN
up to z$\simeq$4 (e.g. Woo \& Urry 2002; McLure \& Dunlop 2004; Warner et
al. 2004; Kollmeier et al. 2006; Vestergaard 2002, 2004; Netzer \&
Trakhtenbrot 2007). The Eddington ratio is then obtained by applying a bolometric
correction to the optical luminosity. These authors seemed to agree on a
practically constant value of  
$\lambda$ ($\lambda$$\simeq$0.25 Kollmeier et al. 2006) at all z, irrespective of luminosity and black hole mass, after selection
effects had been properly taken into account (but see Netzer \&
Trakhtenbrot 2007 for a recent result indicating a dependence of $\lambda$ on both M$_{BH}$ and $z$ up to $z$ $\simeq$0.75).\\

In type 2 AGN the BLR is not visible and the M$_{BH}$ estimate is usually obtained through the empirical
relationships between the black hole mass and the properties of the spheroidal
component of the host galaxy (e.g. Magorrian et al. 1998; Gebhardt et
al. 2000a; Ferrarese $\&$ Merritt 2000; Tremaine et al. 2002). This method has
been applied to a very large sample of type 2 AGN in the local Universe
(Kauffmann et al. 2003; Heckman et al. 2004 here-in-after H04; Kauffmann \& Heckman 2008). At higher $z$, however this
method becomes more difficult to apply, especially in spiral galaxies where it
is necessary to disentangle the contribution of the bulge from that of the
disk. A step forward in this direction has been recently made by  Bian et
al. (2006, here-in-after B06), which were able to estimate the black hole masses for a
subsample of SDSS type 2 AGN at 0.3$<$ $z$ $<$0.83 from the Zakamska et
al. (2003) sample. The black hole mass estimate was obtained from the stellar
velocity dispersion measurements ($\sigma_{\ast}$) and the M$_{BH}$-$\sigma_{\ast}$ relation of Tremaine et
al. (2002).

In type 2 AGN, since the primary optical/UV radiation is much affected
by dust extinction, the bolometric luminosity is generally inferred from the [OIII] emission line luminosity (L$_{OIII}$). 
The [OIII] emission line arises in the much larger narrow line region (NLR),
where the gas is photoionized by the continuum radiation escaping within the opening angle of the torus. The observed flux is therefore
weakly affected by the viewing angle relative to the torus and its luminosity
provides an indication of the nuclear luminosity. The key issue, in order to
estimate $\lambda$ in these AGN, is therefore to have a
bolometric correction factor in hand to be applied to L$_{OIII}$.\\

H04 computed a mean bolometric correction factor to the
observed [OIII] luminosity, and used it to estimate the Eddington ratio
distribution of local (z$\simeq$0.1) type 2 AGN. This bolometric correction factor was also used by B06 to estimate the
$\lambda$-distribution of their sample. B06 found a mean
value of this distribution $\langle \lambda \rangle$ $\simeq$1, which would reveal a population
of high Eddington ratio sources which differs from that observed in the local
Universe and in type 1 AGN. If confirmed, this result would have enormous
impact on the way that massive black holes grow along cosmic time, as most of
this growth as expected to occur at significant redshift and in obscured objects.\\

However, L$_{OIII}$ is only an indirect estimator of the nuclear luminosity,
which depends on the geometry of the system, on the amount of gas, and on any
possible shielding effect which may affect the ionizing radiation seen by the NLR.
This implies an intrinsic scatter in the L$_{OIII}$-L$_{b}$ relation and therefore an uncertainty in the $\lambda$ estimate.  
A further complication arises from the fact that the [OIII] emission line is affected by extinction, likely due to dust within the NLR.
In some cases the reddening due to the NLR can be evaluated, and the extinction-corrected [OIII] luminosity (L$^{c}_{OIII}$) is a more direct estimator
of the nuclear luminosity than L$_{OIII}$. We estimated the [OIII] bolometric
  correction following a similar method adopted by H04 , but we used the extinction-corrected [OIII] luminosity instead of
  the observed one.
 
Then we
used this bolometric correction  to calculate the Eddington ratio of the sources from the B06 sample for which the extinction-corrections are available.
Moreover, we have observed with XMM-{\it Newton} 8 sources of the B06 sample,
in order to have an independent estimate 
of the bolometric luminosity through X-ray luminosity. 

The paper is organised as follows.
In Sect. 2 the bolometric correction estimate is described. Sect. 3 is devoted
to the estimate of the Eddington ratio distribution. Discussion and
conclusions follow in Sect. 4.

\section{The L$_{X}$-L$^{c}_{OIII}$ luminosity relation}


H04  estimated the bolometric
correction (BC) to the observed [OIII] luminosity, L$_{OIII}$, in a two-step process. First they estimated the mean ratio between the monochromatic
continuum luminosity at 5000 $\AA$ and the [OIII]
luminosity, L$_{5000}$/ L$_{OIII}$, in a sample of type 1 AGN. Then they calculated the mean ratio
between the bolometric luminosity and L$_{5000}$ using the average type 1 AGN
intrinsic spectral energy distribution (SED) of Marconi et al. (2004).

They found L$_{b}$/L$_{OIII}$$\simeq$3500. As discussed in H04, 
they did not correct L$_{OIII}$ for dust extinction in the NLR,
because the extinction correction is usually made by measuring the observed Balmer
decrement (H$\alpha$/H$\beta$), and in type 1 AGN this procedure requires deblending the narrow 
components from the broad ones.

In Seyfert galaxies optically classified as type 1.5, 1.8, 1.9 and 2, in
which the narrow components of the H$\alpha$ and
H$\beta$ lines dominate the line profiles (Osterbrock \& Ferland 2006), the estimate of the
Balmer decrement can instead be considered reliable. 
So we used these Seyfert types to estimate the BC to convert the
extinction-corrected [OIII] luminosity, L$^{c}_{OIII}$, to the bolometric
luminosity.

In these Seyfert types the bolometric luminosity
cannot be easily determined because the primary optical/UV radiation is highly obscured.
In X-rays, however, the nuclear emission could be directly estimated,
provided that the absorbing matter is Compton-thin (i.e. $N_H$ $<$
$\sigma_T^{-1}=1.5\times 10^{24}$ cm$^{-2}$; if the matter is Compton-thick,
no nuclear radiation is visible below 10 keV, with the Compton reflection
component usually dominating the spectrum).

Assuming that the X-ray bolometric correction is the same in type 1 and in
type 2 AGN, we used the relation between L$^{c}_{OIII}$ and the absorption
corrected X-ray luminosity and the X-ray
bolometric correction of type 1 and Compton-thin type 2 
AGN to estimate the [OIII] bolometric correction (C$_{OIII}$).\\

\begin{figure}[h!]
\begin{center}
\includegraphics[width=8 cm]{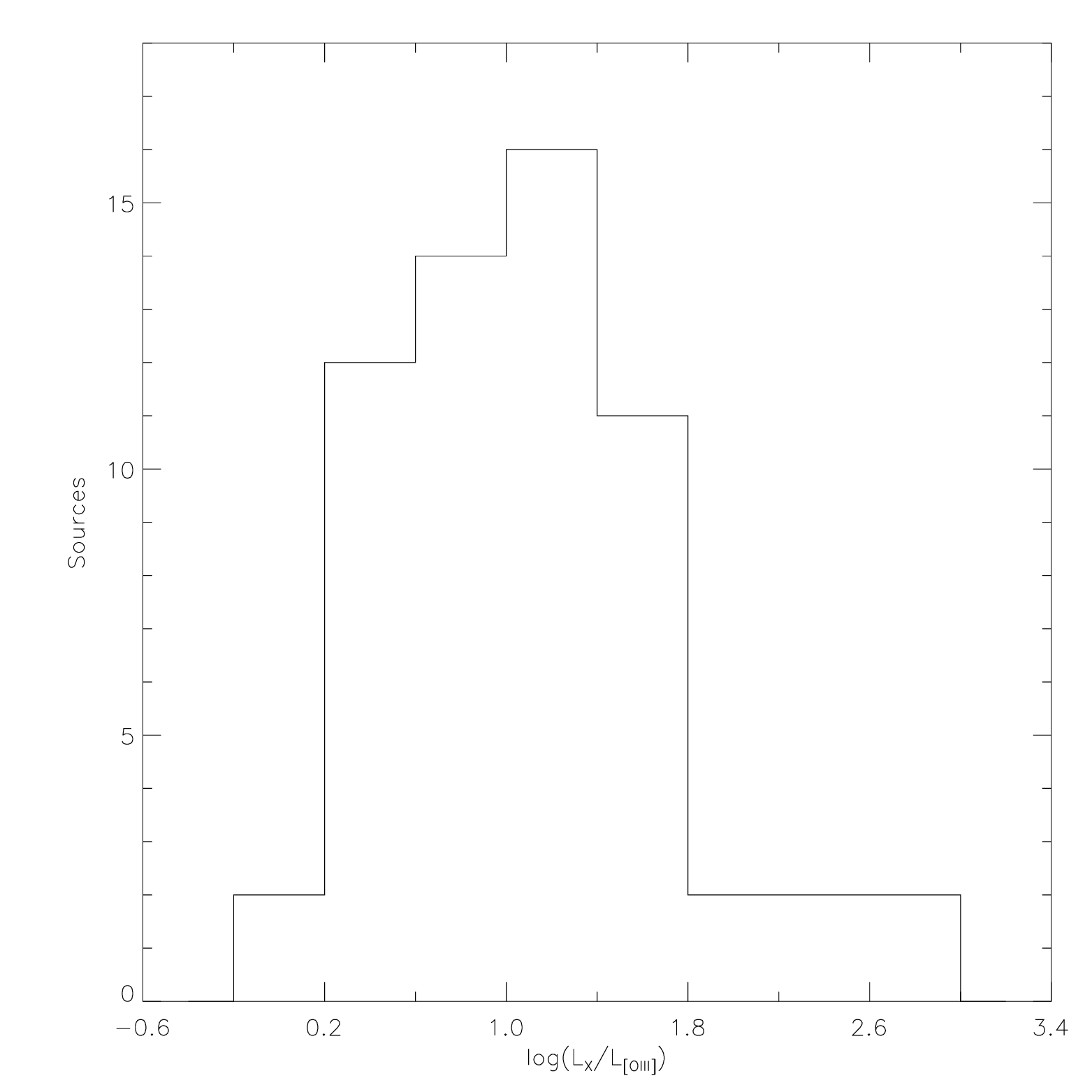}
\includegraphics[width=8 cm]{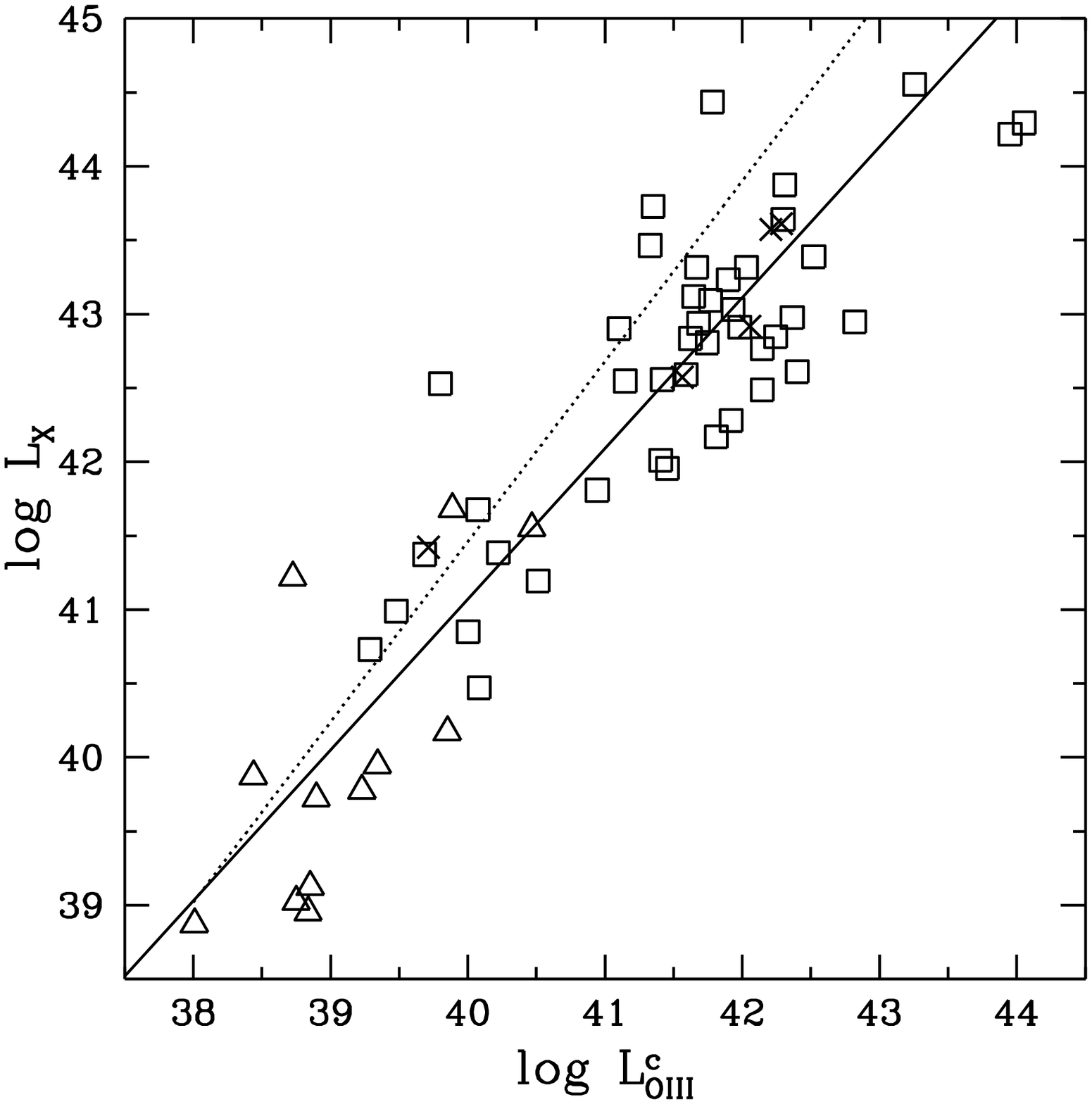}
\caption{Distribution of log(L$_{X}$/L$^c_{OIII}$) (top) and log L$_{X}$-log
  L$^{c}_{OIII}$ relation (bottom). The solid line is
  the best fit relation Eq. (\ref{fit}) for the 61 sources from Mulchaey et
  al. (1994) (crosses) , B99 (squares) and P06 (triangles), while the dotted
  line is the best fit relation obtained by P06.}
\label{oiiix}
\end{center} 
\end{figure}

We have collected a sample from the literature for which reliable estimates  of the Balmer decrement, L$_{OIII}$ and  L$_{X}$ were available. We have discarded Seyfert galaxies with optical classification $\le$ 1.2 and X-ray Compton-thick sources. 
The final sample consists of 61 sources: 5 from Mulchaey et al. (1994)\footnote{We have used L$_{OIII}$ and Balmer decrement from  Mulchaey et al. (1994) and the X-ray
luminosities from Bianchi et al. 2009a.}, 12 from Panessa et al. (2006, P06
hereafter), and 44 from Bassani et al. (1999, B99 hereafter, 8 of these 44 sources
are also included in the P06 sample).
We used the B99 relation to derive  L$^{c}_{OIII}$ from the Balmer decrement:\begin{equation}\label{bassani} 
L^{c}_{OIII}=L_{OIII}\left(\frac{(H\alpha/H\beta)_{obs}}{3.0}\right)^{2.94}.
\end{equation}
which assumes an intrinsic Balmer decrement equal to 3.0 as expected in the NLR
(see Osterbrock \& Ferland 2006).\\
In Fig. \ref{oiiix} (top panel) the distribution of the ratio between the
(2-10) keV luminosity and the extinction-corrected [OIII] luminosity for our sample is shown. The mean of this distribution is 
\begin{equation} \label{pan}
\langle \log \left( \frac{L_{X}}{L^{c}_{OIII}} \right)\rangle=1.09 
\end{equation}
and the dispersion is 0.63 dex.

The bottom panel of Fig. \ref{oiiix} shows L$_{X}$ versus L$^c_{OIII}$. The
correlation is highly significant\footnote{We consider statistically
  significant the correlations whose null hypothesis probability, P, is lower
  than 1$\times$10$^{-3}$, which correspond to a confidence level of
  99.9$\%$.} (Spearman rank correlation coefficient $\rho$=0.86,
P=1.6$\times$10$^{-18}$). In order to quantify any distance-driven
  effects which may be present in a luminosity-luminosity correlation, we also
  performed a ``scrambling test'' (see e.g. Bregman 2005; Merloni et al. 2006;
  Bianchi et al. 2009b). The correlation appears not to be affected by distance effects:
  not even a data-set out of the 100000 simulated ones has a correlation
  coefficient as large as the one measured in the real data-set.

An ordinary least-square bisector procedure gives
the following best fit:
\begin{equation} \label{fit}
\log\left(\frac{L_{X}}{10^{42}erg/s}\right)=(1.11\pm0.10)+(1.02\pm0.06)\log\left(\frac{L^{c}_{OIII}}{10^{42}erg/s}\right)
\end{equation} 
i.e. we found a linear correlation between L$_{X}$ and L$^{c}_{OIII}$ .\\ 
P06 investigated the same correlation finding that  L$_{X}$$\propto$L$^c_{OIII}$$^{(1.22\pm0.06)}$. 
They included in their analysis type 1 AGN, type 2 AGN and 
Compton-thick sources (in the latter case they assumed that the intrinsic X-ray
luminosity is 60 times larger than the observed one) and covered about five orders
of magnitude in  both [OIII] luminosity and X-ray luminosity. 
It should  be recalled that the P06 sample is optically
  selected while our sample, mainly based on the B99 catalogue, includes both optically selected and
  X-ray selected AGN. As discussed in several papers, most recently in Heckman et
  al. 2005 and Netzer et al. (2006, N06 hereafter), X-ray selected samples are likely to be biased towards X-ray bright
  sources and hence produce a larger X/[OIII] ratio compared to
  optically selected ones. The 44 sources which we have selected from the B99
  catalogue have all log L$^{c}_{OIII}$ $>$39.3. The presence of X-ray selected
  objects among these sources should therefore tend to make the
  L$_{X}$-L$^{c}_{OIII}$ relationship steeper than found by P06, which, as it
  can also be seen in Fig. \ref{oiiix}, is exactly the opposite of what we found.\\
  On the other hand, the high luminosity ranges in P06 are populated nearly completely by type 1 AGN,  for
which we consider unreliable the estimate of L$^c_{OIII}$. The steeper slope
of the L$_{X}$-L$^{c}_{OIII}$ relation found by P06  could indeed 
be due to an underestimate of the reddening in these AGN types.\\

N06 studied the relation between L$_{OIII}$
and L$_{X}$  in samples of type 1 and type 2 AGN. They found that the ratio
between L$_{OIII}$ and L$_{X}$ decreases with L$_{X}$, and for type 2 AGN the
same result was obtained also using the extinction-corrected [OIII]
luminosity. In the latter case, they used the B99 catalogue excluding sources
with optical classification $\le$ 1.5 and Compton-thick sources. They found
log(L$^{c}_{OIII}$/L$_{X}$)=(15.0$\pm$4.0)-(0.38$\pm$0.09)log L$_{X}$, which
implies L$_{X}$ $\propto$ L$^c_{OIII}$$^{1.61}$, i.e. a correlation with an even steeper slope
than that obtained by P06.
Our analysis differs from that of N06 only for the inclusion of the AGN
optically classified as type 1.5. Excluding these sources from our sample, we found
for the remaining 52 sources:
$\log(L_{X}/10^{42}erg/s)=(1.08\pm0.11)+(1.04\pm0.06)\log(L^{c}_{OIII}/10^{42}erg/s)$ ($\rho$=0.85,
p=2.0$\times$10$^{-15}$ ), i.e. a  significant linear correlation also in this
case. The discrepancy between the results may be due to the smaller range of
luminosities covered by N06, which do not include sources with L$_{X}$ $\le$ 10$^{40}$erg/s. Unfortunately we
cannot make a clear comparison with N06 because they did not perform a simple
L$_{X}$-L$^{c}_{OIII}$ fit and an extrapolation of the resulting slope from their fit may be
misleading. Moreover N06 did not explicitly report the method they used to compute the
linear regression and the significance of the correlation they found.\\

The relation between L$_{X}$ and L$^c_{OIII}$ can be combined with
  the X-ray bolometric correction (XBC) to obtain the [OIII] bolometric correction factor,
  C$_{OIII}$. Adopting the luminosity-dependent XBC of Marconi et al. (2004), we found a
  mean value of C$_{OIII}$  in the luminosity ranges log L$_{OIII}$=38-40,
  40-42, and 42-44 of  87, 142 and 454 respectively.

  As a cautionary remark, it must be noted that the XBC may suffer from large
  uncertainties.
  
  Marconi et al. (2004) and  Hopkins et al.  (2007) accounted for variations
  of AGN SEDs with luminosity using the anti-correlation between the optical-to-X-ray spectral index 
  ($\alpha_{OX}$) and the luminosity at 2500 $\AA$ (L$_{2500}$) (e.g. Vignali
  et al. 2003a), but
   Hopkins et al. (2007) found that intrinsic spread in AGN SEDs could give
   rise to variation of a factor of $\sim$ 2 in the XBC, even when luminosity
   dependence is taken into account.\\ 

Moreover, the very dependence on luminosity of the X-ray bolometric correction 
has been questioned by Vasudevan \& Fabian (2007), who estimate it from the observed SEDs of 54 AGNs.
They found evidence of a large scatter in the XBC,
with no simple dependence on luminosity, and suggested instead the Eddington ratio 
as the physical quantity  the bolometric correction depends on. \\

Very recently, Kauffmann \& Heckman (2008) anticipated an estimate of C$_{OIII}$.

They found C$_{OIII}$ in the range $\sim$ 500 to 800. Because the details of their
analysis have not been published yet, for the time being we are not able
to discuss the origin of this discrepancy.

\section{The Eddington ratio distribution of type 2 SDSS AGN at 0.3$<z<$0.4}

B06 measured the stellar velocity dispersion, $\sigma_{\ast}$, for 30 sources
of the 0.3$<z<$0.8 SDSS type 2 AGN sample of Zakamska et al. (2003). The
stellar velocity dispersion was obtained by fitting the profile of heavy
element absorption lines which are mainly associated with the old stellar
population in the bulge. Then, they estimated the black hole masses from
$\sigma_{\ast}$, adopting the Tremaine et al. (2002) relation.\\

The B06 subsample is representative of the total
sample of  Zakamska et al. (2003) with respect to L$_{OIII}$. Indeed, the observation of significant stellar absorption features, which is the key point to accurately measure $\sigma_{\ast}$, does not appear to depend on the nuclear luminosity of the sources, as confirmed by a t-test (see B06).

For 15 sources B06 calculated
the [OIII] line luminosity corrected for extinction using the observed Balmer
decrement and Eq. (\ref{bassani}). Therefore all these sources have  $z$ $\leq$ 0.4, because of the limited
wavelength coverage of the SDSS spectroscopy ($\simeq$3800-9200 $\AA$). 
The ratio between L$^{c}_{OIII}$ and L$_{OIII}$, is between 0.6 and 17.8. In three objects the observed Balmer decrement
is lower than the theoretical value, most likely due to an imperfect
subtraction of the starlight. 
We assigned zero extinction to these objects.

Figure 4 in B06 shows the distribution of the Eddington ratios of the 15
sources where we have a reliable estimate of L$^{c}_{OIII}$. This distribution is surprising
for its mean value, $\langle \lambda \rangle$ $\simeq$ 1, because it would reveal a population of
high Eddington ratio sources which is not observed in the local Universe, such
Eddington ratios are neither observed in type 1 AGN at higher redshift.

However, it is important to note that B06 obtained this distribution estimating the bolometric luminosity from L$^{c}_{OIII}$ and the BC given by H04. This bolometric correction applies to L$_{OIII}$ and not to L$^{c}_{OIII}$, therefore B06 overestimated in a systematic manner L$_{b}$ and hence $\lambda$.\\

In this section we used C$_{OIII}$, as computed in Sect. 2, to calculate the Eddington ratio distribution of the
type 2 SDSS AGN of the B06 sample.

\begin{figure}[h!]
\begin{center}
\includegraphics[width=6.5 cm]{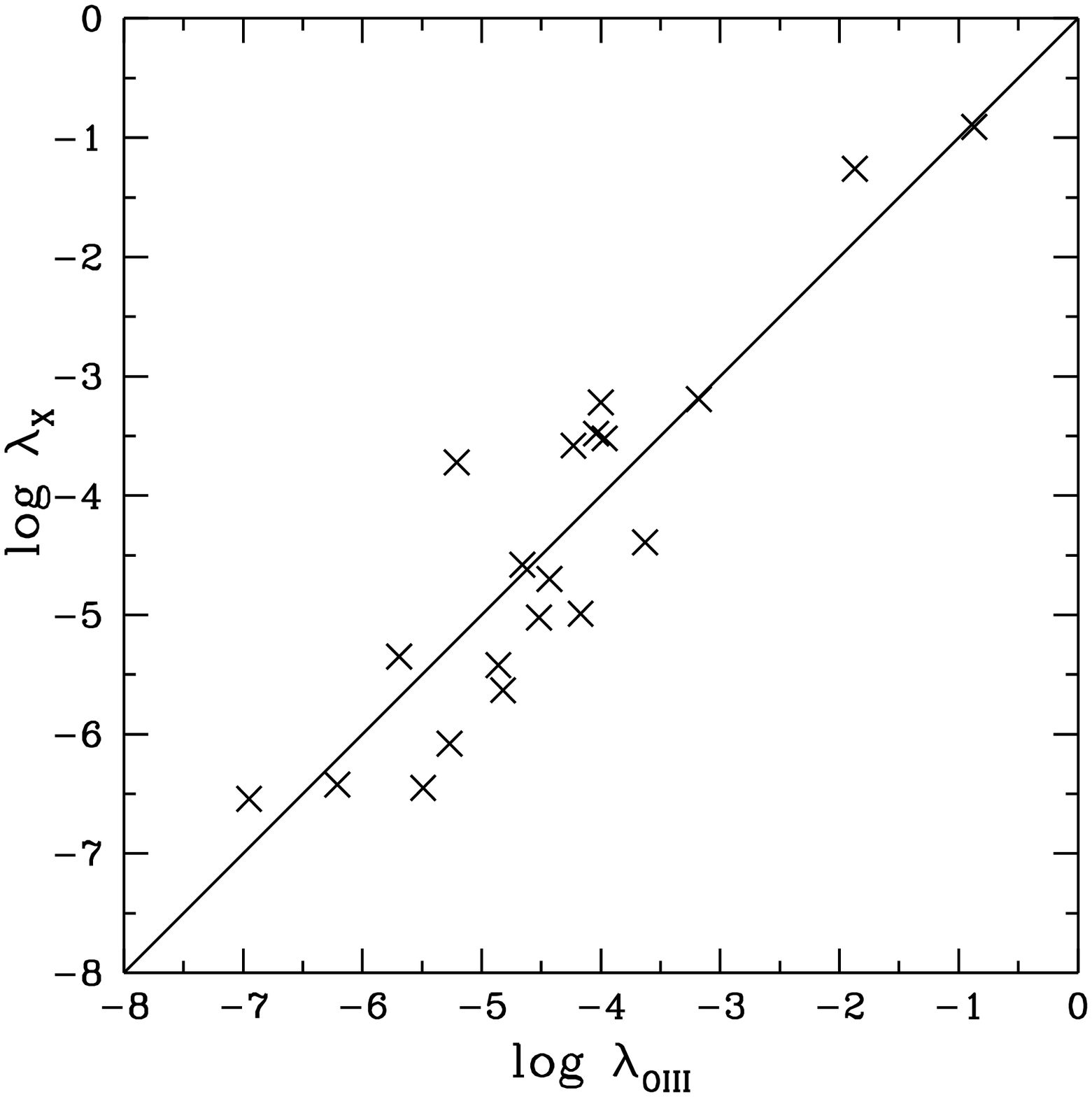}
\includegraphics[width=6.5 cm]{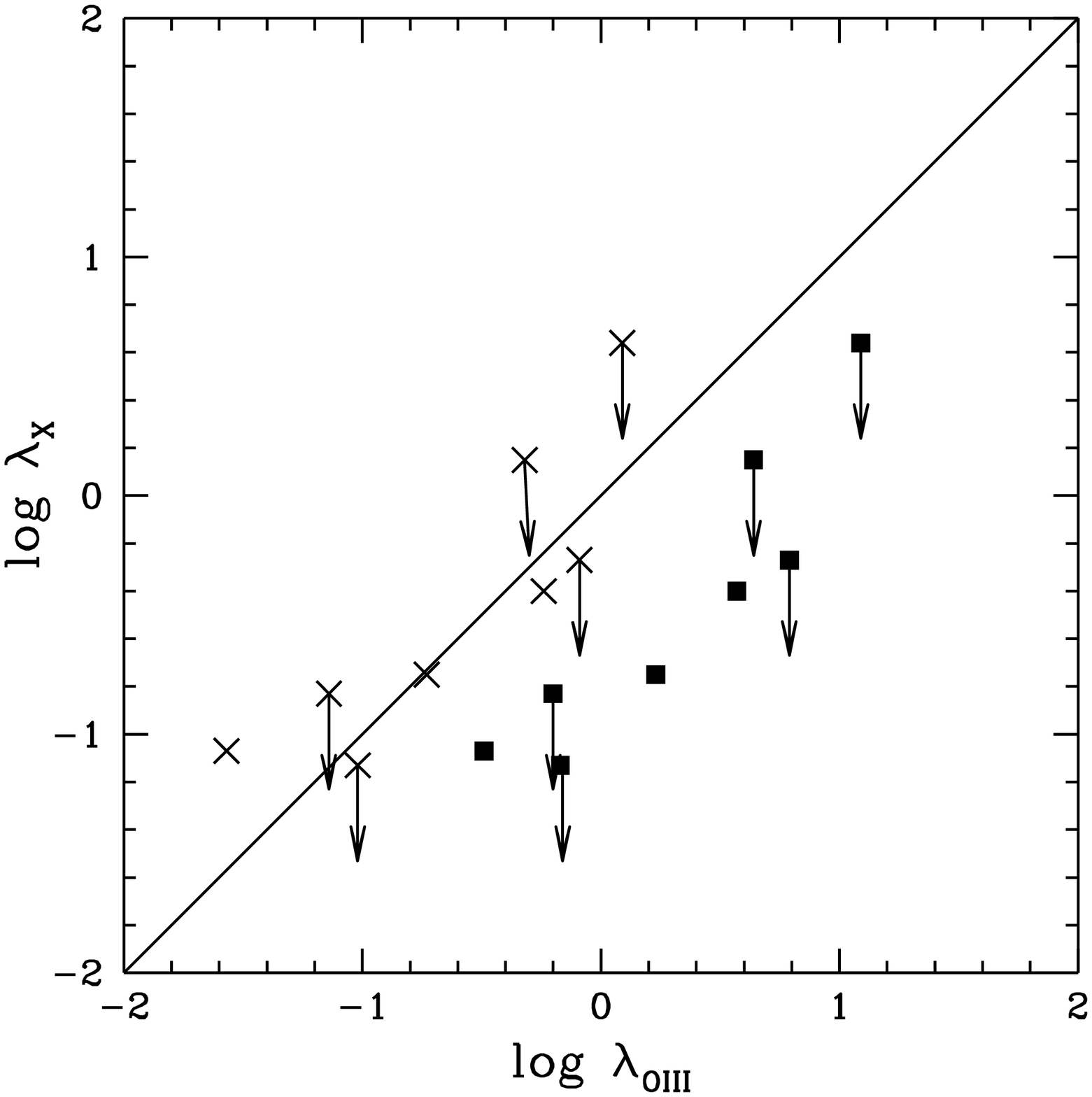}
\caption{Top: $\lambda_{X}$ vs. $\lambda_{OIII}$ for the sources in the P06
  subsample. Bottom: $\lambda_{X}$ vs. $\lambda_{OIII}$ obtained by  B06 (squares) and from
L$^{c}_{OIII}$ and C$_{OIII}$ (crosses), for the B06 sources
  with the XMM-{\it Newton} observations.}

\label{lambda}
\end{center} 
\end{figure}

First of all we have tested
the reliability of deriving $\lambda$ from the [OIII] line luminosity in an
optically selected sample, adopting the C$_{OIII}$ we derived in a collection
of X-ray selected and optically selected sources.
To this aim we used the P06 subsample used in Sect. 2. 
In the top panel of Fig. \ref{lambda} the Eddington ratio estimates obtained from L$^{c}_{OIII}$ 
and C$_{OIII}$ ($\lambda_{OIII}$) are compared with those obtained from L$_{X}$ and the XBC of Marconi et
al. 2004 ($\lambda_{X}$) (the black hole masses are taken from Table 2 of
P06). We found a good agreement between the X-ray and the [OIII] estimates.\\
For the 8 SDSS AGN of the B06 sample for which we have 
XMM-{\it Newton} observations we can repeat the above check.
The bottom panel of Fig. \ref{lambda} shows $\lambda_{X}$ versus $\lambda_{OIII}$ obtained from L$^{c}_{OIII}$ 
and C$_{OIII}$ (crosses) and versus $\lambda_{OIII}$ obtained by B06 (squares). 
As expected, the $\lambda_{OIII}$ obtained by B06 are systematically
larger than those derived from the X-ray luminosity, while this effect
disappears when the proper bolometric correction is used.

The black hole mass, the [OIII] line luminosity, the extinction-corrected [OIII] line luminosity,  the (2-10) keV observed flux (F$^{obs}_{X}$) and the intrinsic (i.e. de-absorbed) (2-10) keV luminosity (L$^{int}_{X}$) of the X-ray observed sources are listed in Table \ref{SDSS_source}. The X-ray data reduction and the estimate of L$^{int}_{X}$ either from X-ray spectral fitting or extrapolated from the observed X-ray count rates (or upper limits) are reported in the Appendix.\\

\begin{figure}[h!]
\begin{center}
\includegraphics[width=7 cm]{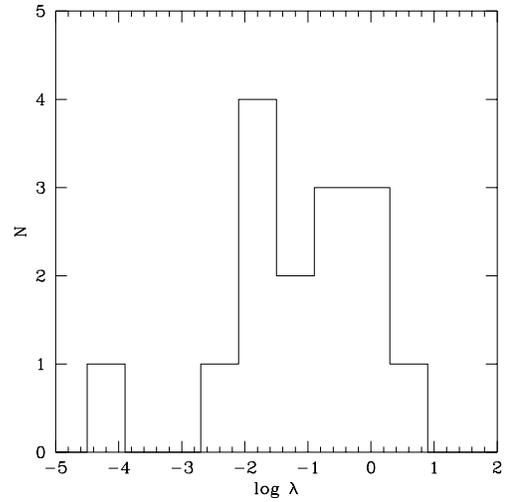}
\caption{Eddington ratio distribution of the B06 sample which is obtained estimating $\lambda$ from L$^{c}_{OIII}$ and C$_{OIII}$.}
\label{histo_Bian}
\end{center} 
\end{figure}

Figure \ref{histo_Bian} shows the $\lambda$-distribution of the B06 sample
obtained using L$^{c}_{OIII}$ and C$_{OIII}$ to estimate the bolometric
luminosity. This distribution is obviously shifted to lower $\lambda$ values
with respect to that obtained by B06: the mean of this distribution is
$\langle \lambda \rangle$$\simeq$0.1, a value which is similar to that
estimated in other AGN samples.\\

We have checked whether the width of the observed $\lambda$-distribution
indicates a physical spread in the Eddington ratios or it can be due
to the dispersion in the L$^{c}_{OIII}$- L$_{b}$ relation. In practice, 
we checked if the observed distribution is consistent with a log-normal
distribution centered at log$\lambda$=-1 and with width equal to the
dispersion in the L$_{X}$-L$^{c}_{OIII}$ relation (0.63 dex). A 
Kolmogorov-Smirnov test gives D=0.27, which
corresponds to a probability of 0.18 for the null hypothesis that the two
distributions are drawn from the same parent population.

\begin{table*}
\begin{tabular}{|c|c|c|c|c|c|}
\hline
Source & M$_{BH}$ &L$_{OIII}$ &L$^{c}_{OIII}$ & F$^{obs}_{X}$& L$^{int}_{X}$ \\
(1)&(2)&(3)&(4)&(5)&(6)\\
\hline
SDSSJ075920.21+351903.4& 8.36 & 7.59&8.84&1.3$\times 10^{-14}$ &43.92$^{\ast}$   \\
SDSSJ083945.98+384319.0& 8.16 & 8.60$^{\dagger}$&9.71$^{\dagger}$ &2.3$\times 10^{-13}$ & 44.26    \\
SDSSJ084041.05+383819.9& 7.71 & 8.62& 9.48&$<$4.4$\times 10^{-14}$&$<$44.07   \\
SDSSJ092318.04+010144.7&8.53&8.94&9.31&$<$4.0$\times 10^{-14}$&$<$44.24\\ 
SDSSJ094820.38+582526.4& 7.01 & 7.89&9.08&$<$4.2$\times 10^{-14}$&$<$44.17   \\
SDSSJ133735.01-012815.6& 8.75 & 8.72&9.56&$<$6.0$\times 10^{-14}$&$<$44.25   \\
SDSSJ143027.66-005614.8& 7.61 & 8.36&9.23&$<$8.6$\times 10^{-14}$&$<$44.38   \\
SDSSJ143928.23+001538.0& 8.01 & 8.08&9.22&6.6$\times 10^{-14}$& 43.88\\
\hline
\end{tabular}
\caption{(1) SDSS name. (2) log of the black hole mass in units of M$_{\odot}$ from B06. (3) log of the [OIII] line luminosity in units of L$_{\odot}$ from Zakamska et al (2003). (4) log of the extinction-corrected [OIII] line luminosity in units of L$_{\odot}$ from B06. (5) Observed X-ray flux in units of erg s$^{-1}$ cm$^{-2}$(see the Appendix).  (6) log of the de-absorbed X-ray luminosity in units of erg s$^{-1}$ (see the Appendix).
($\ast$) Compton-thick source: L$^{int}_{X}$  is obtained assuming that the
ratio between the reflected and the primary component is  0.05.
  ($^{\dagger}$)
 this source has $z$=0.424 so the SDSS spectrum does not
  cover the spectral range necessary to measure the H$\alpha$ line. The [OIII] line luminosity and the Balmer decrement measurements are obtained from observation collected with the  2.2 m telescope at the Centro Astron\'{o}mico Hispano Alem\'{a}n (CAHA) at
Calar Alto. }

  \label{SDSS_source}
\end{table*}

\section{Conclusions}

In this paper we estimate the bolometric correction
factor, C$_{OIII}$, needed to convert the extinction-corrected [OIII] line luminosity to
bolometric luminosity. 

We determined C$_{OIII}$ in a two-step process. First we studied the
L$_{X}$-L$^{c}_{OIII}$ relation in a sample of 61 Seyfert galaxies with
reliable measurement of both L$^{c}_{OIII}$ and L$_{X}$, i.e. we selected
galaxies with reliable measurement of the Balmer decrement
(Seyfert galaxies with optical classification $\ge$1.5) and X-ray Compton-thin
sources. The bolometric correction C$_{OIII}$ was then obtained by combining the mean
ratio between L$_{X}$ and L$^{c}_{OIII}$ with the luminosity-dependent X-ray bolometric
correction of Marconi et al. (2004). We found a
  mean value of C$_{OIII}$  in the luminosity ranges log L$_{OIII}$=38-40,
  40-42, and 42-44 of  87, 142 and 454 respectively.\\
 Similarly, H04 estimated the [OIII] bolometric correction combining
the mean ratio between L$_{5000}$ and L$_{OIII}$  in a sample of
type 1 AGN with the 5000 Angstrom continuum to L$_{bol}$ correction from Marconi et
al. (2004). H04 did not correct L$_{OIII}$ for the extinction in the NLR,
therefore their bolometric correction applies to the ``observed'' [OIII] luminosities.
Compared to H04 our estimate of the [OIII] bolometric correction is not
subject to the scatter due to the extinction in the NLR, which we estimated
to be about 0.6 dex from our sample.\\

The [OIII] line luminosity is an indirect estimator of the nuclear
luminosity, depending on the geometry of the system, and on any
possible shielding effect which may affect the ionizing radiation seen by the NLR.  
The amount of gas which is exposed to the ionizing radiation depends on the
torus opening angle. Bianchi et al. (2007, and references therein) confirmed the 
Iwasawa-Taniguchi (IT) effect, i.e. the anti-correlation between the iron line equivalent
width and the X-ray luminosity. The simplest explanation for the IT effect is in terms
of a decrease of the torus covering factor (and hence an increase of its opening angle)
with luminosity, which may also explain the observed decrease with
luminosity of the ratio between mid-IR and bolometric luminosities (Maiolino et al. 
2007; Treister et al. 2008). If this explanation is correct, we would expect a less
than linear relation between L$_{X}$ and L$^{c}_{OIII}$, differently from what we
found (the situation is even worse with the more than linear relations found by
P06 and N06). This problem will be addressed in a forthcoming paper.

We used C$_{OIII}$ to estimate the Eddington ratio distribution of
type 2 SDSS AGN at 0.3$<z<$0.4 from the B06 sample. B06 found that the mean of
this distribution 
is $\langle \lambda \rangle$ $\simeq$ 1, thus revealing a population of high Eddington ratio
sources which is absent in the local Universe and in type 1 AGN at
higher redshift (e.g. Woo \& Urry 2002; McLure \& Dunlop 2004; Warner et
al. 2004; Kollmeier et al. 2006; Vestergaard 2002, 2004; Netzer \&
Trakhtenbrot 2007). 
However, we have demonstrated,  with the help of the X-ray
luminosities, that B06 have overestimated in a
systematic manner the bolometric luminosities and hence the Eddington
ratios, which we found to be
$\langle \lambda \rangle$ $\simeq$0.1, similar to what found in other AGN samples.

\section*{Acknowledgements}
This work benefited from an Italy-Spain Integrated Actions, reference HI-0079.
AL, GM, and SB acknowledge financial support from ASI (grant I/088/06/0).
XB and FJC acknowledge financial support by the Spanish Ministerio de Ciencia e Innovaci\'on under project ESP2006-13608-C02-01
. The authors thank Cristian Vignali for useful discussions, and the anonymous
referee for his constructive comments.
Based on observations obtained with {\it XMM-Newton}, an ESA science mission
with instruments and contributions directly funded by ESA Member States and
the USA (NASA). Based on observation collected with the
2.2 m telescope at the Centro Astron\'{o}mico Hispano Alem\'{a}n (CAHA) at
Calar Alto, operated jointly by the Max-Planck Institut f\"ur Astronomie and the Instituto de Astrof\'{i}sica de Andaluc\'{i}a (CSIC).

\appendix

\section{The X-ray data}

Of the 8 sources presented in this paper, 7 were observed as a part of our
XMM-{\it Newton} proposal ID 050206 and ID 055120, and one was observed serendipitously.\\
The data have been processed using the XMM-{\it Newton} Science Analysis Software (SAS) v.8.0. The raw event files (the Observation Data Files, ODF), have been linearized with the XMM-SAS pipeline EPPROC for the PN camera, and we generated calibration files using the XMM-SAS pipeline CIFBUILD. Event files were cleaned from bad pixel (hot pixels and events out of the field of view) and we have selected events spread at most in two contiguous pixels (pattern=0-4).
We removed periods of high background levels by analysing the light curves of the count rate at energies higher than 10 keV. Response functions (the ancillary response file and the redistribution matrix file) for spectral fitting were generated using the SAS tasks RMFGEN and ARFGEN. 
We have extracted the PN counts in a circular region, with radius of 15$^{''}$, centered at the source position, and the background counts in the source-free regions close to the target. 

The SDSS name, redshift, exposure times (filtered for good time intervals), X-ray PN count rates (or 5$\sigma$ upper limits) in the rest frame energy ranges (0.5-2) keV and (2-10) keV are listed in Table \ref{xraycount}.\\ 

Of the 8 sources, two are detected (i.e. the Poisson probability of a spurious background excess is less than 2.7$\times$10$^{-3}$) in the (0.5-10) keV band, two in the
soft band (0.5-2 keV) only  and one in the hard band (2-10 keV) only.

\begin{table*}
\begin{tabular}{|c|c|c|c|c|}
\hline
Source &z &Exp.&Count rate (0.5-2) keV   & Count rate (2-10) keV \\
(1)&(2)&(3)&(4)&(5)\\

\hline
SDSSJ075920.21+351903.4&0.328&19.16&  $<$1.7 &0.9$^{+0.4}_{-0.3}$  \\ 
\hline
SDSSJ083945.98+384319.0&0.424&16.00&  1.1$^{+0.3}_{-0.3}$ &6.7$^{+0.7}_{-0.7}$ \\
\hline
SDSSJ084041.05+383819.9&0.313&16.00&  $<$2.3&$<$2.0  \\ 
\hline
SDSSJ092318.04+010144.7&0.386&23.36&  1.2$^{+0.4}_{-0.3}$ & $<$1.8\\
\hline
SDSSJ094820.38+582526.4&0.353&31.38&  1.6$^{+0.3}_{-0.3}$ & $<$1.9\\
\hline
SDSSJ133735.01-012815.6&0.329&6.11&   $<$4.2 &$<$2.7\\
\hline
SDSSJ143027.66-005614.8&0.318&7.49&   $<$4.0 & $<$3.9 \\ 
\hline
SDSSJ143928.23+001538.0&0.339&19.32&  1.2$^{+0.4}_{-0.3}$ &3.3$^{+0.5}_{-0.4}$  \\ 
\hline
\end{tabular}
\caption{(1) SDSS name. (2) Redshift. (3) Exposure time filtered for good time intervals in units of Ks. (4) Observed count rates in rest-frame (0.5-2) keV band in units of 10$^{-3}$ counts s$^{-1}$. (5) Observed count rates in rest-frame (2-10) keV band in units of 10$^{-3}$ counts s$^{-1}$ }
\label{xraycount}
\end{table*}

Sufficient counts for a spectral fitting in the 2-10 keV band were collected
only from three sources: SDSSJ075920.21+351903.4  SDSSJ143928.23+001538.0 and SDSSJ083945.98+384319.0.   

\begin{figure}[h!]
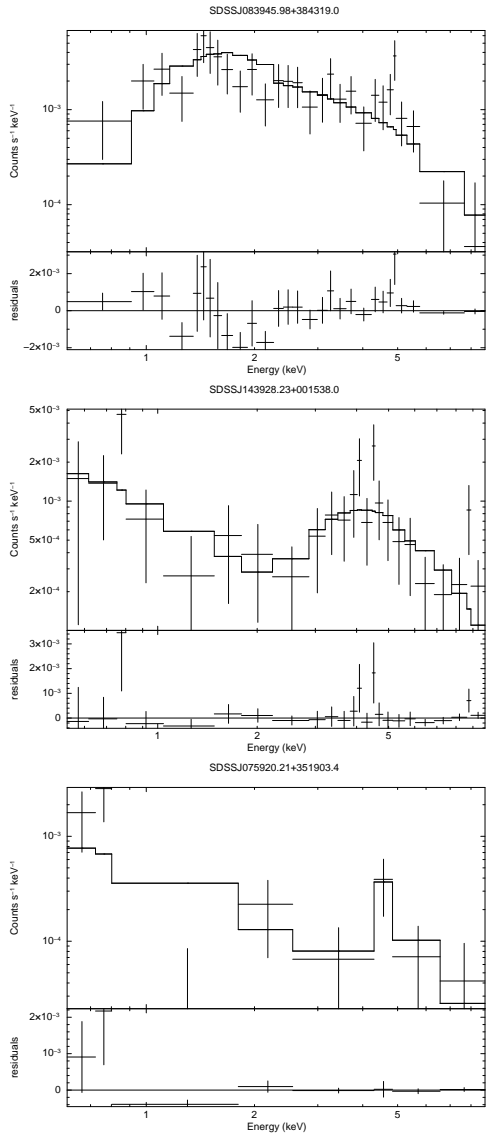

\begin{center}
\includegraphics[width=5 cm,angle=270]{SDSS083945.ps}
\includegraphics[width=5 cm,angle=270]{SDSS143928.ps}
\includegraphics[width=5 cm,angle=270]{SDSS075920.ps}
\caption{XMM-{\it Newton} spectra of SDSSJ083945.98+384319.0,
  SDSSJ143928.23+001538.0 and SDSSJ075920.21+351903.4 from top to bottom.}
\label{spectra}
\end{center} 
\end{figure}

Spectral analysis was carried out with XSPEC v.12.4.0 using data accumulated
in energy bins with 5 counts each and the C-statistic (Cash 1979).
The spectral fits are shown in Fig. \ref{spectra}.
\begin{table*}
\begin{tabular}{|c|c|c|c|c|}
\hline
Source &$\Gamma$ &N$_H$& F$^{obs}_{2-10}$ &log L$^{int}_{X}$\\
(1)&(2)&(3)&(4)&(5)\\
\hline
SDSSJ075920.21+351903.4& 1.9 (fixed)&--&1.3$\times 10^{-14}$ &43.92$^{\ast}$\\
SDSSJ083945.98+384319.0&1.9 (fixed)&3.28$^{+0.95}_{-0.78}$ $\times10^{22}$&2.3$\times 10^{-13}$ & 44.26  \\
SDSSJ143928.23+001538.0&1.9 (fixed)&27.1$^{+15}_{-10.2}$ $\times 10^{22}$&6.6$\times 10^{-14}$& 43.88\\
\hline
\end{tabular}
\caption{(1) SDSS name. (2) Power-law photon index. (3) Hydrogen column density  (4) observed flux in the (2-10) keV band in units of erg s$^{-1}$ cm$^{-2}$. (5) log of the absorption-corrected luminosity in the (2-10) keV band in units of erg s$^{-1}$. ($\ast$) L$^{int}_{X}$  is obtained assuming that the ratio between the reflected and the primary component is  0.05}
\label{fit_par}
\end{table*}

The X-ray spectra in the (0.5-10) keV band  are fitted with an absorbed power-law continuum plus a Compton-reflection component and a soft-excess component, modelled with a power-law with photon index equal to the one of the primary continuum. The parameters of the fits along with the observed (2-10) keV flux, and the de-absorbed (2-10) keV luminosity are shown in Table \ref{fit_par}. In SDSSJ075920.21+351903.4 an iron line with equivalent with of about 2.6 keV is also detected, revealing that this source is Compton-thick. In Compton-thick sources the primary X-ray emission is totally suppressed below 10 keV, the hard X-ray emission is accounted for by reflection of the primary nuclear emission off the inner walls of optically thick circumnuclear matter.

The luminosity reported in Table \ref{fit_par} is obtained assuming that the ratio between the reflected and the primary component is  0.05, similar to what observed in the Circinus galaxy, one of the best studied Compton-thick sources (Matt et al. 1999).

The upper limits in the X-ray luminosity, listed in Table \ref{SDSS_source},
for the other sources are extrapolated from the observed count rates (column
(5) of Table \ref{xraycount}) assuming a typical spectrum of Compton-thick
sources. The spectrum is assumed to be a power-law with photon index
$\Gamma$=1.9 absorbed by column density of N$_H$=1.5$\times10^{24}$, plus a
pure Compton reflection component with an iron line (at 6.4 keV and with
equivalent width EW=1 keV with respect of the pexrav component). The ratio
between the reflected and the primary component is assumed to be 0.05. We have
chosen this spectrum because we wanted to study the extreme case of
obscuration which implies the higher intrinsic X-ray luminosities and
therefore the higher bolometric luminosities.

We have used this method to estimate L$^{int}_{X}$ also in the sources detected only in the soft band, because from the comparison
between the observed X-ray luminosity and that predicted from L$^{c}_{OIII}$ and Eq. (\ref{pan}) we argued that they are heavily absorbed sources. Therefore the soft X-ray emission is not the primary soft X-ray emissions but a soft excess component.

\end{document}